\documentclass[twocolumn,showpacs,amsmath,amssymb]{revtex4-1}
\usepackage{graphicx}

\begin{document}

\title{A 6D interferometric inertial isolation system}
\author{C.~M.~Mow-Lowry}
\affiliation{School of Physics and Astronomy and Institute of Gravitational Wave Astronomy, 
University of Birmingham, Edgbaston, Birmingham B15 2TT, United Kingdom}
\author{D.~Martynov}
\affiliation{LIGO, Massachusetts Institute of Technology, Cambridge, Massachusetts 02139, USA}
\affiliation{School of Physics and Astronomy and Institute of Gravitational Wave Astronomy, 
University of Birmingham, Edgbaston, Birmingham B15 2TT, United Kingdom}
\date{\today}

\begin{abstract}
We present a novel inertial-isolation scheme based on six degree-of-freedom (6D) interferometric sensing of a single reference mass. It is capable of reducing inertial motion by more than two orders of magnitude at 100\,mHz compared with what is achievable with state-of-the-art seismometers. This will enable substantial improvements in the low-frequency sensitivity of gravitational-wave detectors. The scheme is inherently two-stage, the reference mass is softly suspended within the platform to be isolated, which is itself suspended from the ground. The platform is held constant relative to the reference mass and this closed-loop control effectively transfers the low acceleration-noise of the reference mass to the platform.  A high loop gain also reduces non-linear couplings and dynamic range requirements in the soft-suspension mechanics and the interferometric sensing. 
\end{abstract}

\pacs{04.80.Nn, 07.10.Fq}
\keywords{gravitational wave detectors; seismic isolation; inertial sensors}
\maketitle

%%%%%%%%%%%%%%%%%%%%%%%%%%%%% Introduction %%%%%%%%%%%%%%%%%%%%%%%%%%%%%%

Gravitational waves (GW) from black hole and neutron star binaries have recently been observed by the LIGO and Virgo detectors~\cite{first_detection, second_detection, GW170814, GW170817}. These interferometers were designed to be sensitive to gravitational waves in the frequency range from 10\,Hz up to a few kHz. The low frequency band ($<30$\,Hz) is particularly important for studying intermediate mass black holes, with masses of $\sim 10^3 M_\odot$, and for accumulating signal-to-noise ratio from lighter sources. However, during the first science runs of the Advanced LIGO and VIRGO network the signal was mostly accumulated between 30 and 300\,Hz, and the observed black hole masses were from 7 to 70$\,M_\odot$. Despite the sophistication and successful operation of LIGO's internal seismic isolation systems~\cite{isi_matichard, isi_part_1, isi_part_2}, the sensitivity of the detector degrades at low frequencies due to non-stationary control noises~\cite{den_nb}.

If no new technology is developed, future GW observatories will face similar problems. A recent paper (and supplemental material) analysed the critical noise-couplings and the upgrades required to allow astrophysically interesting sensitivity at frequencies below 10\,Hz~\cite{Yu5Hz}. The conclusion is that the scientific returns are substantial, supported by studies of future facilities such as the Einstein Telescope~\cite{ET_2012} and Cosmic Explorer~\cite{lungo_2017}. The noise performance for new technologies required to achieve this sensitivity are calculated. In this paper we analyse a 6D interferometric inertial isolation system that can enable gravitational-wave observations at 10\,Hz and below.

Although it is designed for GW detectors, the calculated inertial isolation performance of the proposed isolation system is directly applicable to other instruments. The predicted inertial translation is substantially below that of seismically-stable bedrock, to a level of approximately 1\,nm/$\sqrt{\rm Hz}$ at 0.1\,Hz. This is required for torsion-bar gravitational-antennae~\cite{Ando2010, McManus2016, Shimoda2018}, and can reduce aliased-noise and Doppler-shifting in atom-interferometer instruments such as MIGA~\cite{Canuel2017MIGA}. 
A range of other laboratory experiments can also benefit from such a low acceleration environment, from space-borne accelerometer development~\cite{Zhou2010}, to measurements of small forces such as inverse-square-law tests~\cite{Adelberger2003}.

% In precision experiments, such as LIGO, the core optics are mounted on or suspended from a  platform that is actively or passively isolated from ground motion. Suppression of seismic noise between 10\,mHz and 10\,Hz is particularly important for GW detectors in order to avoid coupling control noise into the GW channel~\cite{Yu5Hz}. In order to achieve this goal, we propose to use a single reference mass that is softly suspended in all six degrees of freedom from an isolated platform. Its position relative to the platform is monitored using six interferometers.

% Figure \ref{refMass} shows the design concept: control forces are applied to the platform with high gain to stabilize the relative position, effectively transferring the inertial stability of the reference mass to the platform. This process is similar to the drag-free control of satellites, where thrusters hold the spacecraft a constant distance from a freely-floating reference mass that is shielded from external forces. Crucially, all degrees of freedom are simultaneously isolated in a low-noise fashion, all but eliminating typical cross-coupling problems. 

\begin{figure}[b]
  \begin{center}
  \includegraphics[width=8.6cm]{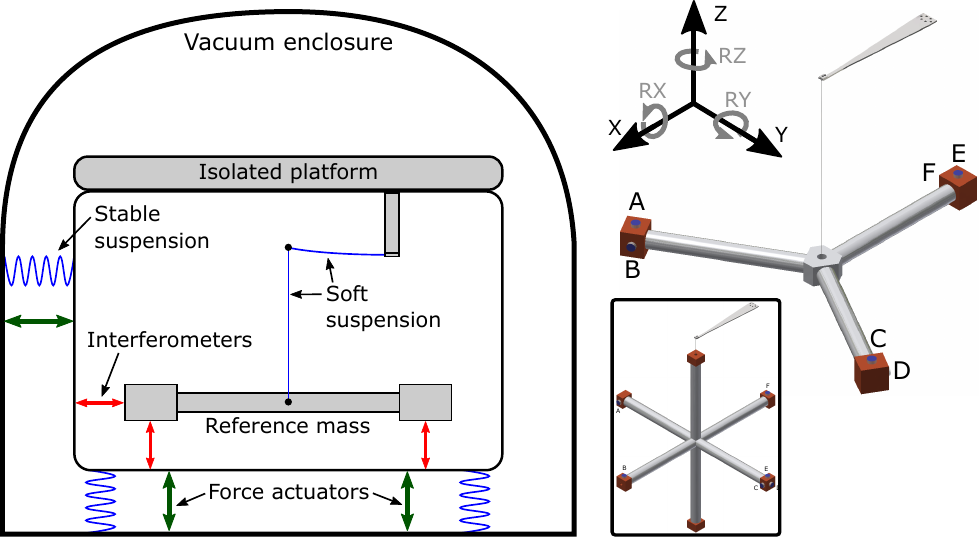}
  \end{center}
  \caption{A 2-d representation of the isolation architecture (left) and a design concept for the reference mass and suspension (right). Letters indicate interferometric sensing locations. Inset right: an alternative configuration with equal moments of inertia in the three principal axes that reduces Newtonian noise in RX and RY at the expense of size and complexity.} 
  \label{refMass}
\end{figure}

The novel design concept, shown in Figure \ref{refMass}, addresses four of the most challenging problems in terrestrial low-frequency inertial isolation: sensing and force noise of the reference mass, Tilt-to-horizontal coupling, cross-coupling, and dynamic range. This is achieved by, first, employing low-noise interferometric sensors, an ultra-high-quality suspension system, and a high-vacuum enclosure. Second, the tilting resonances are pushed to sufficiently low frequencies to allow them to be precisely sensed. Third, all six degrees of freedom are similarly quiet, such that cross-coupling doesn't spoil the performance. Finally, a high-gain control system suppresses actuation noise and dynamic range requirements in both the soft suspension and the sensing interferometers.

The central reference mass is suspended such that the weakest possible restoring forces, limited only by material properties and geometric constraints, are applied in all degrees of freedom. It is sensed by six interferometers and high-gain control holds their outputs constant such that the isolated platform is fixed relative to the reference mass to a resolution limited by sensing noise. This process is comparable to the drag-free control successfully employed in LISA Pathfinder~\cite{lpf2016}, where the spacecraft was manoeuvred around one reference mass, while a second mass was weakly (and periodically) constrained with electro-static actuators. 

A key concept in both our proposed 6D isolator and LISA pathfinder is that all six degrees of freedom are simultaneously low-noise, all but eliminating the cross-coupling. In Pathfinder, however, the aim was to shield the reference masses from external forces. In our design, the high-gain control imposes the inertial stability of the reference mass onto the isolated platform. This inherently two-stage topology produces `terrestrial drag-free control', and it has the potential to transform active inertial isolation.

\section{Low-frequency inertial isolation}  
The primary limitation of low-frequency inertial isolation systems is tilt-to-horizontal coupling~\cite{Lantz09.BSA}. The best active system is currently LIGO's ISI, which measures inertial tilt by subtracting the vertical outputs of horizontally separated sensors such as Trillium T-240 seismometers. The angular inertial performance is approximately the same as the seismometer self-noise, but the horizontal sensitivity is degraded by a factor of $\sqrt{1 + g^2/\omega^4}$, where $g$ is local gravitational acceleration and $\omega$ is angular frequency \cite{Lantz09.BSA}.

The resulting noise prevents LIGO from significantly reducing the RMS motion of the ground at micro-seismic frequencies, $\sim 0.1-0.2$\,Hz. Moreover, the platform motion is amplified between 10 and 70\,mHz. The residual platform motion causes angular fluctuations of the test masses with an RMS amplitude of 0.1\,$\mu$rad. To reduce this motion to less than the divergence angle of the LIGO cavities, $\sim 30$\,nrad, wavefront-sensors~\cite{ifo_asc} with a control bandwidth of a few Hz are required~\cite{den_thesis}. The self-noise of these wavefront-sensors is $~10^{-14}$\,rad$/\sqrt{\rm Hz}$, and the angular controls degrade LIGO's sensitivity all the way up to 20-30\,Hz~\cite{den_nb}.

Two approaches are currently being explored in the LIGO-Virgo collaboration to reduce low-frequency tilt-coupling. The first aims to develop a seismometer that is insensitive to tilt in a particular frequency band~\cite{sei_tilt_free_2015, sei_tilt_free_2016}. The second aims to actively stabilize the tilt-motion of the isolated platforms using 1D rotation sensors. Custom-built devices with considerably lower self-noise than commercial seismometers have already been installed outside the vacuum system at LIGO~\cite{sei_uwash_tiltmeter_2014, sei_uwash_tiltmeter_2017}.

%The solution proposed here is to use a single reference mass that is softly suspended in all six degrees of freedom from an isolated platform. Its position relative to the platform is monitored using six interferometers. Figure \ref{refMass} shows the design concept: control forces are applied to the platform with high gain to stabilize the relative position, effectively transferring the inertial stability of the reference mass to the platform. This process is similar to the drag-free control of satellites, where thrusters hold the spacecraft a constant distance from a freely-floating reference mass that is shielded from external forces. 

Our proposed 6D system has two clear advantages over other approaches; it requires only one device to stabilize a platform in all degrees of freedom, and all mechanical degrees of freedom are weakly constrained. In contrast, LIGO's isolation scheme is over-constrained, with at least 12 sensors on each stage, and tilt-free seismometers and 1D rotation sensors both have some stiff degrees of freedom that make them more prone to cross-coupling. 

Suspending and tuning a single mass that is very soft in all three angular degrees of freedom will present significant experimental challenges. We will exploit the overlap with 1D rotation sensors and torsion-bar antennae - optically sensed large-moment reference masses - and apply similar solutions for dealing with large long-term drifts and centre-of-mass tuning.

%These improvements will increase the duty cycle of LIGO and allow the observation of gravitational waves at 10\,Hz. A more detailed study of the effect on LIGO performance, when coupled with some additional upgrades, is contained in~\cite{Yu5Hz}. This isolation scheme also opens a way towards even lower observational frequencies by future gravitational wave detectors such as Einstein Telescope~\cite{ET_2012} and Cosmic Explorer~\cite{lungo_2017} that have not yet provided a solution to low-frequency technical noise couplings.

%%%%%%%%%%%%%%%%%%%%%%%%%%%%% Mechanics %%%%%%%%%%%%%%%%%%%%%%%%%%%%%%

\section{Mechanical design of the reference mass}
The key feature of the mechanical design is the suspension of a large-moment reference mass from a single fused-silica suspension wire. The combination of low-loss, high-strength material, and a large-moment reference mass is required to reduce the rotational thermal noise enough to allow for a transformational reduction in inertial translation at 0.1\,Hz. The large tilt-measurement baseline and high precision interferometric sensing are also required to improve the resolution of the rotation readout.

We propose to suspend the reference mass from a single quasi-monolithic fused-silica fibre~\cite{Cumming2012.CQG, Aisa16.NIMA}, similar to the LIGO and Virgo test mass suspensions, which implies that the fibre attachments do not significantly increase the thermal noise of the mass. High performance in the tilting degrees of freedom (RX and RY as seen in Fig.~\ref{refMass}) is crucial. The tilt-modes must have very low resonant frequencies to distinguish them from translational motion and reduce sensing noise. 

In our design these criteria are met by combining a large reference mass, a thin, highly-stressed fused silica suspension, and interferometric sensing. The three-spoked design in Fig.~\ref{refMass} meets these requirements.

\renewcommand{\arraystretch}{1.4}
\setlength{\tabcolsep}{1em}
\begin{table}[]
\centering
\begin{tabular}{ccc}
\textbf{Resonances} & \multicolumn{1}{c}{\textbf{Frequency}} & \textbf{Q} \\ 
\hline
X, Y                     & 0.5\,Hz              & 10$^9$            \\
Z                        & 1\,Hz                & 10$^3$            \\
RX, RY                   & 5\,mHz               & 10$^5$            \\
RZ                       & 0.7\,mHz             & 10$^6$            \\
\end{tabular}
\caption{Fundamental resonant frequencies and quality factors of the reference mass suspension.}
\label{tab:resonances}
\end{table}

Weakly constrained mechanical oscillators are prone to long-term drift. To keep the reference mass at its operating point and actively damp the resonant modes, low-frequency control forces must be applied. The magnitude of the required force is dependent on residual stresses and creep in the suspension materials. To prevent the applied forces from spoiling the isolation performance, we anticipate applying a combination of low-force capacitive actuation, with a dynamic range of $\sim$10$^6$, and step-wise re-adjustment, either manually or via slip-stick actuation.

%Stuck here to enhance float-placement
\begin{figure*}
 \includegraphics[width=0.45\linewidth]{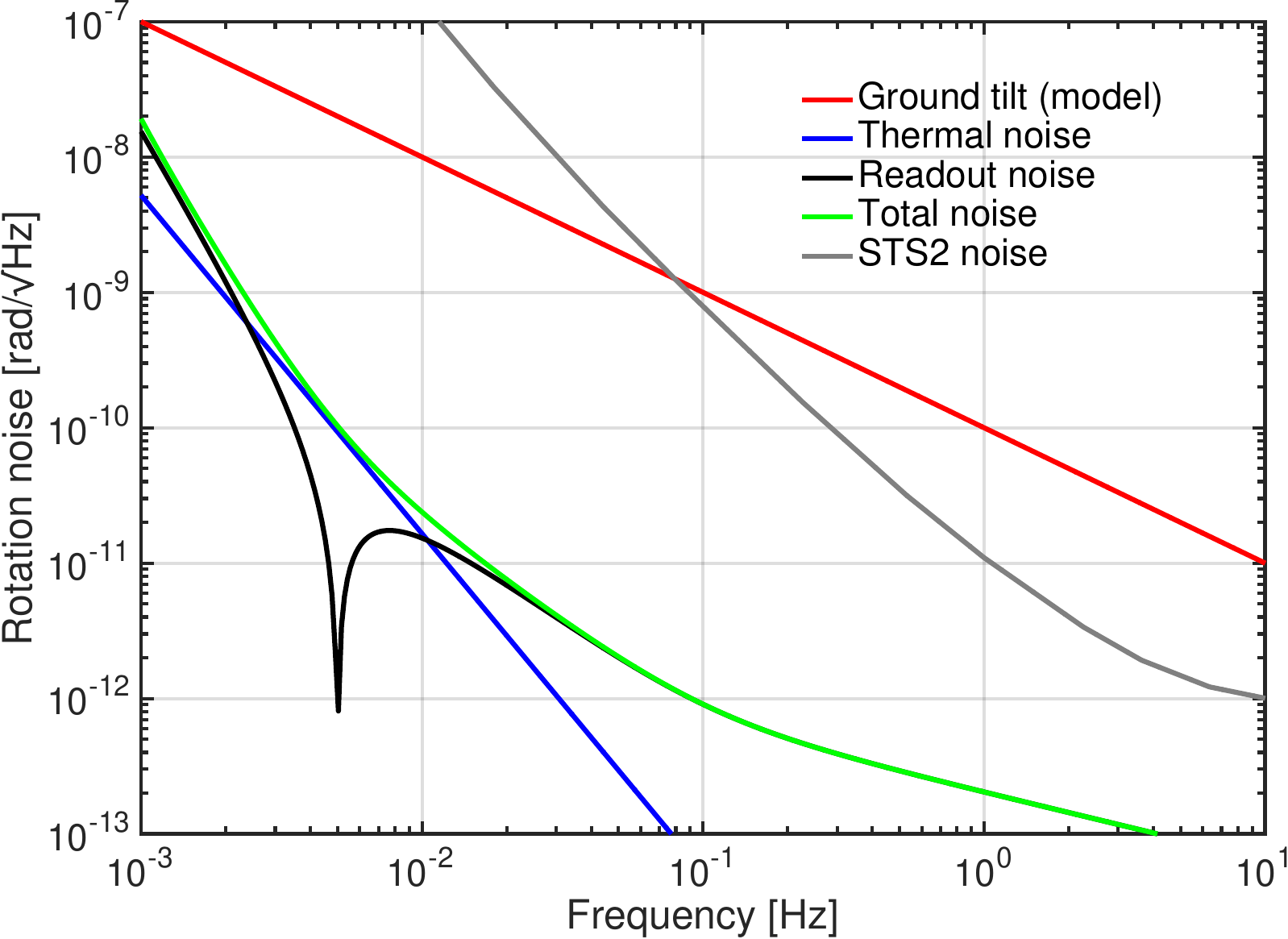}
\quad\quad
 \includegraphics[width=0.45\linewidth]{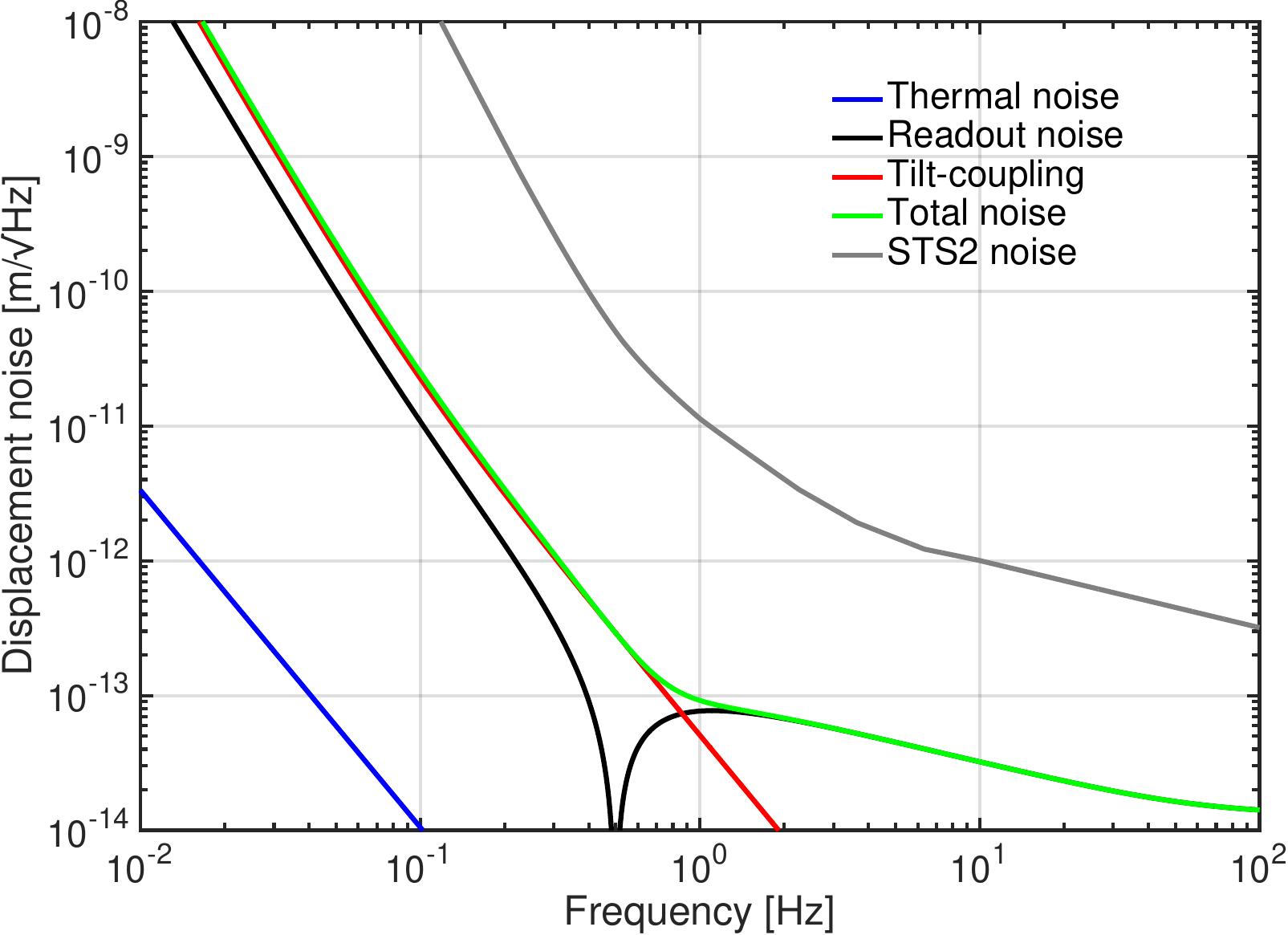}
\caption{A noise budget showing (left) the predicted angular self-noise for rotation around the horizontal axes of the 6D isolator and (right) the predicted horizontal displacement self-noise assuming that the angular noise couples with a factor of $g/\omega^2$. }
\label{fig:nb}
\end{figure*} 

In the vertical (Z) direction a standard metal blade-spring provides compliance and supports the total mass load \cite{Beccaria1998}. It might be possible to employ a lower-loss material, but in keeping with other conservative assumptions, we assume that a maraging steel blade will be used, with $\sim$1000 times larger loss than fused silica.

The fundamental resonant frequencies and quality factors of the six principle modes are shown in Table~\ref{tab:resonances}. The horizontal translational (X and Y) degrees of freedom are essentially pendulum modes. The vertical (Z) resonance is dominated by the elastic compliance of the blade spring, and the torsion (RZ) by the shear restoring torque of the fibre. The material loss angle of the fibre, $\phi_{\rm mat}$, dominated by surface and thermo-elastic loss, is conservatively assumed to be $10^{-6}$~\cite{Heptonstall14.CQG}. A fuller description of the assumed material properties and suspension stiffnesses is given in the supplemental material.

The tilt-mode stiffnesses are affected by both elastic and gravitational restoring terms. Assuming the elastic bending length is much shorter than the pendulum length, the elastic angular stiffness of the bending fibre, $\kappa_{\rm el}$, can be calculated as shown in~\cite{Cagnoli00.pla} (and more generally as shown in\cite{Speake17.Met}) resulting in
\begin{align}
\kappa_{\rm el} & = \frac{1}{2} \sqrt{m g E I_{\rm a}},
\label{eq:kel}
\end{align}
where $mg$ is the tension in the fibre, $E$ is the elastic modulus, and $I_{\rm a}$ is the second moment of area, given by $I_{\rm a} = \frac{\pi}{4} r^4$ for a circular cross-section fibre of radius $r$.

For the parameters chosen here, the elastic resonant frequency in tilt is $\omega_{\rm el} = 2 \pi \times 23$\,mHz. The inertial-equivalent sensing noise rises with $1/\omega^2$ below this frequency, as seen in Fig.~\ref{fig:nb} (left). The proposed resonant frequency of $\omega_{\rm RX} = 2 \pi \times 5\,$mHz was chosen to be low enough such that sensing noise in tilt does not degrade translational performance near the blending frequency of $\sim$10\,mHz. Therefore, the centre of rotation should be between 180 and 190\,$\mu$m below the centre of mass, resulting in a resonant frequency between 4 and 7\,mHz. Tuning of the centre of mass relative to the centre of rotation with micron-level precision has been demonstrated, with stable long-term performance, in 1D rotation sensors with comparable mass and moment of inertia \cite{sei_uwash_tiltmeter_2014}. 

%%%%%%%%%%%%%%%%%%%%%%%%%%%%% Sensitivity %%%%%%%%%%%%%%%%%%%%%%%%%%%%%%

% Here we need some modified text: The reference mass is completely unconstrained (at relevant frequencies), and the caging structure is uniquely constrained to it. The core benefit is that there are no large forces applied to the reference mass at all. All degrees of freedom are similarly quiet, so cross-coupling is essentially irrelevant. Gain heirarchy can be used to ensure the tilt degrees are more tightly constrained. Some Z-coupling is assumed due to imperfect readout and offset angles modifying the equations of motion.

\section{Sensitivity analysis} 
We compare the performance of our 6D isolation system with the best commercial seismometers, showing that a substantial reduction of platform motion at LIGO is possible, in turn reducing the required bandwidth of auxiliary control loops by a factor of approximately 5. 
The predicted performance of the isolation system is limited by several noise sources, and we present an analysis of the dominant noise terms, including contributions from thermal noise of the suspension, sensing noise, temperature gradients, control noise, cross-coupling, and input ground motion. 

Since tilt-to-horizontal coupling is the dominant factor in low-frequency horizontal performance, Fig.~\ref{fig:nb} shows a noise analysis of both tilt (RX, RY) and horizontal (X, Y) degrees of freedom. The dominant noise source at nearly all frequencies is the sensing noise of the proposed interferometers. The key noise terms are described below.

% Ground rotation
The input ground rotation must be strongly suppressed by closed-loop control. The `ground rotation' trace in Fig.~\ref{refMass} (left) is a crude fit based on measurements made at the LIGO observatories, with a typical $1/\omega$ slope and a value of 1\,nrad$/\sqrt{\rm Hz}$ at 0.1\,Hz. An aggressive blending frequency is used to provide strong isolation at 10\,mHz. 

% Readout noise
The reference mass is sensed using 3 horizontal and 3 vertical Michelson-type interferometers at locations marked with capital letters in Fig.~\ref{refMass} (right). We assume that the interferometers are of the kind, and with sensitivity, recently reported in~\cite{Cooper18.CQG}. The noise is dominated by the analogue input noise of the ADC system above 0.1\,Hz, and by temperature and air-pressure fluctuations at lower frequencies. 
%$10^{-13} \times \frac{1}{\sqrt{f}} \,{\rm m/\sqrt{Hz}}$ is assumed, and this has recently been demonstrated at frequencies between 0.1 and 10\,Hz in
Significant sensitivity gains can be expected through a combination of environmental and electronic improvements before reaching shot noise, which is $\sim 10^{-15} \,{\rm m/\sqrt{Hz}}$ for a few milliwatts of power. In the analysis shown here a stick-figure fit is made to the reported sensitivity curve assuming no improvements are made.

% Thermal noise
The power spectral density of the thermal-noise torque applied by the bottom hinge to the reference mass is determined by the real part of the mechanical impedance
\begin{align}
S_{\rm torq}(\omega) & = 4 k_{\rm B} T {\rm Re}(Z(\omega)) = 4 k_{\rm B} T \phi_{\rm eff} \kappa_{\rm el}/\omega,
\label{eq:thTorque}
\end{align}
where $k_{\rm B}$ is Boltzmann's constant, $T$ is the temperature in kelvin, and $\phi_{\rm eff}$ is the effective loss angle of the oscillator. The relation between the effective loss angle and the material loss angle is dependent on the geometry of the spring, and in this case $\phi_{\rm eff} = \phi_{\rm mat}/2$~\cite{Cagnoli00.pla}. The supplemental material contains some further discussion concerning the thermal-noise torque.

%Since the sensitivity of an inertial isolation system is determined by the free response of the reference mass, the thermal noise trace in Fig.~\ref{fig:nb} (left) is given by
%\begin{align}
%\sqrt{S_{\rm rot}(\omega)} & = \dfrac{\sqrt{S_{\rm torq}(\omega)}}{\omega^2 I_{\rm RX}}.
%\label{eq:thRot}
%\end{align}
%The total suspended mass and the fibre diameter may be adjusted to achieve a constant fibre stress such that $m \propto r^2$. Combining Eqs.~\ref{eq:kel} and \ref{eq:thTorque}, we find that the thermal-noise torque is proportional to $m^{\frac{3}{4}}$, and with Eq.~\ref{eq:thRot}, the resulting angular motion is proportional to $m^{-\frac{1}{4}}$, as long as the effective radius does not change with mass. 
%
%This weak dependence of thermal noise on the suspended mass means that technical considerations can dictate the design, at the factor of a few level, rather than the elastic resonant frequency or the final quality factor of the RX and RY resonances. The size of the mass, $R$, is chosen such that the rotational thermal noise (barely) limits the final longitudinal inertial sensitivity shown in Fig.~\ref{fig:nb}. It can be readily shown that a steel or tungsten wire would have approximately two orders of magnitude worse performance thermal noise performance, and as such unable to produce satisfactory inertial isolation in this 6D system.

All thermal noise curves are calculated assuming that the loss angles of the fibre and blade-spring are frequency independent. We assume that the loss angle of the fibre is greater than the measured loss at the thermo-elastic loss peak~\cite{Heptonstall14.CQG}, allowing some margin to account for the (small) anticipated clamping losses at the metal-glass interfaces. 

% Cross-coupling
By design, our 6D seismometer is quiet in all degrees of freedom - the high-gain control system will reduce motion down to our noise limits below a few hertz. This means that cross-coupling, typically a major problem for soft mechanical systems, is all but eliminated as an issue. Additionally, the force-noise (including actuator noise) and cross-couplings of the `stable isolation' are actively suppressed by the loop gain, and have a negligible impact on the sensitivity.

What remains important is the fundamental tilt-to-horizontal coupling, where we assume that the final inertially-controlled residual rotation shown in Fig.~\ref{fig:nb} (left) couples with $g/\omega^2$ into horizontal translation, and coupling from the (relatively) noisy vertical (Z) direction. Fortunately, the cross-coupling between vertical and other degrees of freedom is small, and we assume a factor of $10^{-3}$ from thermally-driven vertical motion into horizontal and tilt motion, consistent with Advanced LIGO suspension modelling~\cite{Cumming2012.CQG}. Should it prove necessary, the weak electro-static actuators can be used to fine-tune the reference mass alignment to minimize this coupling.

% Thermal expansion
Thermal expansion makes two distant points of the platform move differentially. We stabilize motion at one point of the platform while the optic is suspended from a different location. We estimate the amplitude spectral density of the thermal-expansion-driven motion, $\sqrt{S_{\rm te}}$, of the optic suspension point at 1\,mHz to be
\begin{equation}
	\sqrt{S_{\rm te}} = \alpha \Delta T \Delta L = 10^{-9} \frac{\Delta T}{1\,{\rm mK}}\frac{\Delta L}{10\,{\rm cm}}
	\frac{\rm m}{\rm \sqrt{Hz}},
\end{equation}
where $\alpha \approx 10^{-5}$ is the coefficient of thermal expansion, $\Delta T$ is the temperature fluctuation, and $\Delta L$ is the distance between the optic suspension point and the seismometer. The temperature gradients that cause this relative motion decrease in magnitude as $1/f$ and they are additionally low-pass filtered by LIGO's vacuum enclosure with a timescale of $\sim$5 hours. Therefore, above 1\,mHz this noise decreases as $1/f^2$ and should not limit isolation performance.

During the sensitivity analysis, we have attempted to employ conservative performance estimates where possible. The quality factor of fused silica suspension fibres has been observed in all-glass and glass-metal systems at the level indicated. Cross-coupling from the noisy vertical (Z) into the crucial tilt (RX, RY) degrees of freedom  is assumed at a reasonable level, even though there is no expected linear coupling. The interferometric sensing noise performance has already been demonstrated, and improvements in that performance can be expected at low frequencies. 

Existing small-force (and torque) torsion balance experiments have shown residual forces to be below the requirements of this proposed experiment, see for example \cite{Carbone04.CQG}. What remains untested is achieving (nearly) suspension thermal-noise limited performance at very low frequencies with a reference mass with such a large moment of inertia. A detailed technical study combined with experimental demonstration will be required to demonstrate the system's practical feasibility.

%%%%%%%%%%%%%%%%%%%%%%%%%%%%% Impact %%%%%%%%%%%%%%%%%%%%%%%%%%%%%%

\section{Isolation performance and impact}
\begin{figure}
 \includegraphics[width=0.95\linewidth]{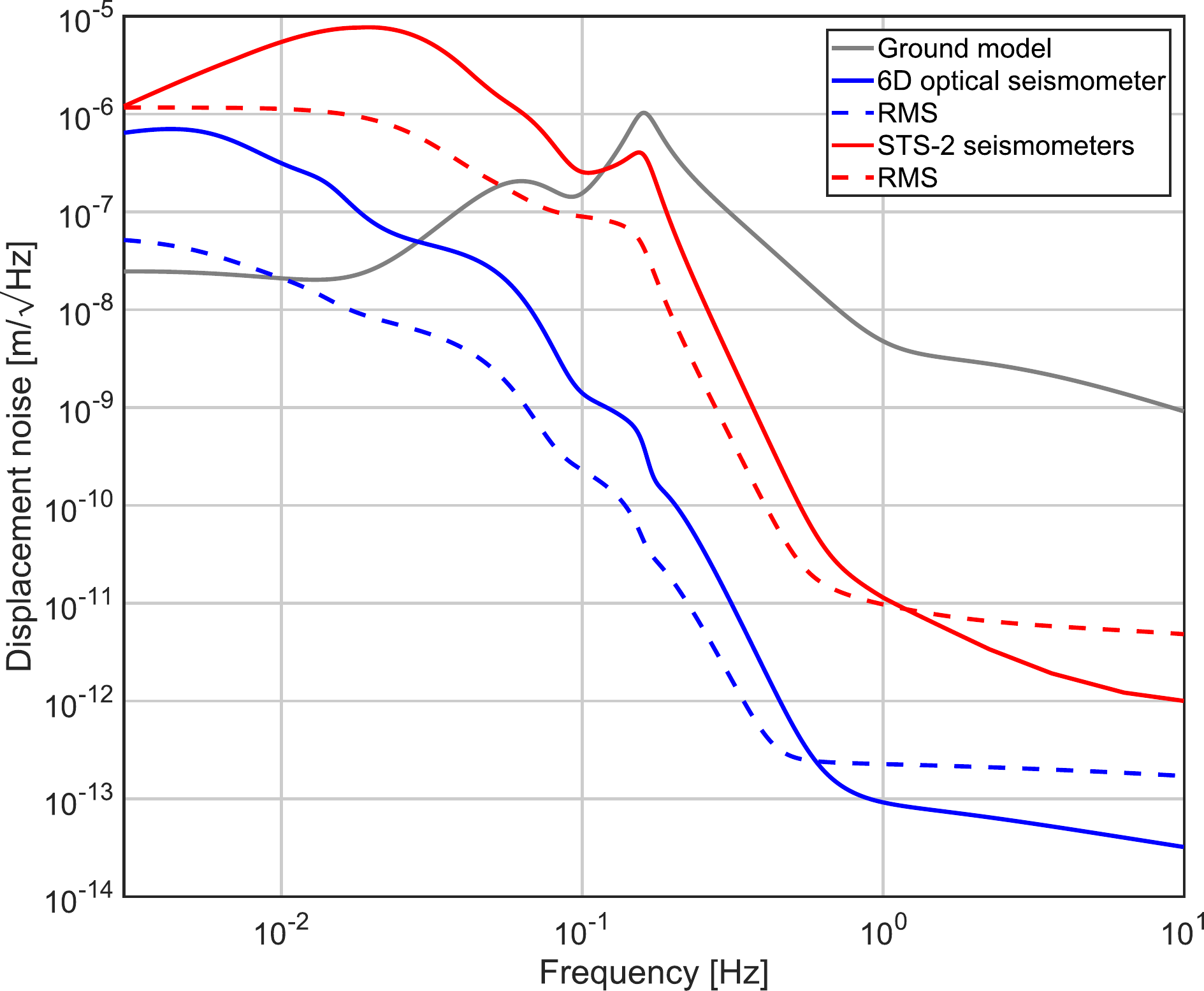}
\caption{A comparison of the achievable residual displacement comparing the use of STS-2 seismometers and the proposed 6D isolation system.}
\label{fig:isi_motion}
\end{figure}

The `6D optical seismometer' trace in Fig.~\ref{fig:isi_motion} represents the total inertial motion in the horizontal direction when all control loops are closed. To avoid adding low-frequency inertial sensing noise to the platform, the feedback signals are blended with position sensor signals at $\sim$10\,mHz, effectively `locking' the platform to the ground at very low frequencies. At higher frequencies, ground motion leaks through the blending filters (shown in the supplemental material), reducing performance near 0.1\,Hz. 

The predicted performance shown is more than two orders of magnitude better than what is possible with state of the art STS-2 seismometers. The RMS displacement is significantly less than a nanometre for all frequencies above 100\,mHz. Such isolation will drastically simplify the lock acquisition procedure of gravitational wave detectors, which currently suffer from large low-frequency motion of $\sim$100\,nm over 100\,seconds. 

During the first science runs, the coincident duty cycle of the two Advanced LIGO instruments was 44\%, and each interferometer was in `observation mode' 66\% of the time. For 18\% of the run, each instrument was not observing due to ground motion at micro-seismic frequencies, wind, and (small) earthquakes. Assuming that the 6D isolation system enables operation in the presence of this elevated motion, the individual instrument duty cycle could be 81\%, in turn increasing the coincident observation time to 66\%.

Noise from auxiliary degrees of freedom currently limits the performance of Advanced LIGO performance below 20-30\,Hz~\cite{den_nb}. The dominant contribution is from angular control of the test masses, where a closed-loop bandwidth of 2.5\,Hz is required to suppress the angular motion to a few nanoradians RMS. With our proposed seismometer, the bandwidth of the auxiliary control loops will be reduced to 0.5\,Hz, rendering control noise from these degrees of freedom insignificant for the gravitational wave readout above 5\,Hz.

Finally, the proposed seismometer opens a way towards detecting GWs at 5\,Hz using ground based detectors. Potential upgrades for the LIGO instruments have recently been proposed~\cite{Yu5Hz}, and a 6D seismometer is a necessary but not sufficient component. A second required element is much improved local sensors for damping of the main suspension chain, and the interferometers proposed here also meet that requirement. The supplemental material in~\cite{Yu5Hz} addresses the performance impact (and requirements) for several proposed upgrades. There is no guarantee that eliminating the low-frequency noise sources that currently limit gravitational-wave detectors will allow astrophysically relevant sensitivity below 10\,Hz, but the kind of isolation presented here is necessary even if it is not sufficient. 

The rich scientific rewards for low-frequency sensitivity, including the detection of intermediate mass black holes up to $2000 M_\odot$, studies at cosmological distances of $z \approx 6$, and the possibility of pre-merger alerts for neutron star coalescence~\cite{Cannon2012}, demand that serious effort is made to address known limitations.

% The 6D interferometric isolation system proposed in this Letter is a foundational element for achieving the required performance at low frequency.

%%% Summary %%% Not included for now.

%\textit{Summary} --- We have proposed a new architecture for inertial isolation and performed a noise analysis demonstrating its feasibility with current technology. If the predicted residual motion is achieved, it will be more than two orders of magnitude lower than what is currently possible, and the benefits for interferometric gravitational-wave detectors will be dramatic.

%%%%%%%%%%%%%%%%%%%%%%%%%%%%% Acknowledgements %%%%%%%%%%%%%%%%%%%%%%%%%%%%%%

\begin{acknowledgments}
The authors wish to thank K. Venkateswara, V. Frolov, G. Hammond, B. Lantz, C.C. Speake, J. Harms, A. Freise, and R. Mittelman for useful discussions and advice. C. M-L has received funding from the European Union’s Horizon 2020 research and innovation programme under the Marie Sklodowska-Curie grant agreement No 701264. D.M. is supported by the Kavli Foundation.

\end{acknowledgments}

\appendix

\section{Equations of motion}
\begin{figure}
  \begin{center}
  \includegraphics[width=7cm]{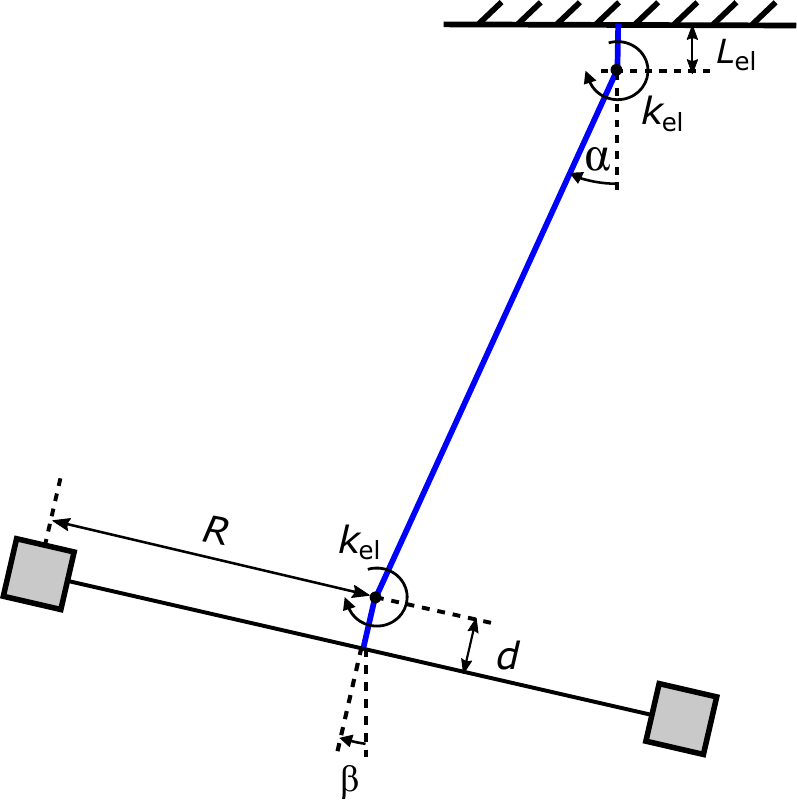}
  \end{center}
  \caption{A simplified schematic for calculating the equations of motion in the coupled degrees of freedom. The pendulum is of length $L$ and the reference mass has a total mass $m$ and a moment of inertia of $I_{\rm RX} \: (\,= I_{\rm RY})$.} 
  \label{pendSchematic}
\end{figure}

%\textit{Equations of motion} --- 
The equations of motion for the reference mass and suspension are simple in Z and RZ, where only the elastic bending of the blade-spring and the torsional compliance of the suspension fibre (respectively) contribute restoring forces. The X and RY (and similarly Y and RX) degrees of freedom are intrinsically coupled. Fig.~\ref{pendSchematic} shows a schematic view of the relevant parameters for determining the Lagrangian of the generalized coordinates $\alpha$ and $\beta$, which are approximately equal to the final Cartesian coordinates $X$ and $RY$ (or $Y$ and $RX$). The design parameters for the reference mass and its suspension are shown in Table \ref{tab:params}. 

\renewcommand{\arraystretch}{1.4}
\setlength{\tabcolsep}{1em}
\begin{table}[]
\centering
\begin{tabular}{lc}
\textbf{Design property}                       & \textbf{Value}              \\ \hline
Suspended mass                                 & 3.8\,kg                     \\
Moments of inertia ($I_{\rm RX}, I_{\rm RY}$)  & 0.36\,kg\,m$^2$             \\
Moment of inertia ($I_{\rm RZ}$)               & 0.72\,kg\,m$^2$             \\
Width ($R$) 		                           & 0.5\,m                      \\
Pendulum length                                & 1\,m                        \\
Vertical spring extension ($\Delta z$) 		   & 0.25\,m \vspace{6pt}        \\ 
\multicolumn{2}{c}{\textbf{Fused-silica suspension fibre properties}}        \\ \hline
Elastic modulus ($E$)                          & 72\,GPa                     \\
Shear modulus ($S$)                            & 15-25\,GPa                  \\
Diameter                 	                   & 200\,$\mu$m                 \\
Fibre stress             	                   & 1200\,MPa                   \\
Material loss angle ($\phi$)                   & $\quad < 10^{-6}$ \vspace{6pt} \\
\end{tabular}
\caption{Design parameters of the reference mass and its suspension.}
\label{tab:params}
\end{table}

The flexing of the suspension fibre provides an elastic restoring torque a distance of $L_{\rm el}$ from its attachment point with an angular spring $\kappa_{\rm el}$. The displacement $d$ between the centre of mass and the lower bending point of the fibre can be adjusted to be positive or negative (through the use of a lockable moving mass) to tune the resonant frequencies of the tilting modes.
The 2-D Lagrangian is
\begin{align}
\mathfrak{L} \approx & \frac{m L^2}{2}\dot{\alpha}^2 + m L d \dot{\alpha} \dot{\beta} + \frac{I_{\rm RY}}{2} \dot{\beta}^2  \nonumber \\
  &  - \frac{m g L}{2} \alpha^2 - \frac{m g d}{2} \beta^2 - \kappa_{\rm el} \alpha^2 + \kappa_{\rm el} \alpha \beta - \frac{ \kappa_{\rm el}}{2} \beta^2,
%\label{eq:lagrangian}
\end{align}
which is valid for $d \ll L, R$ and $\alpha, \beta \ll 1$. 

From the Lagrangian, the (coupled) equations of motion are derived and solved, creating a stiffness matrix. From this the normal-mode frequencies can be determined as a function of the design parameters, although in practice the design height $d$ is adjusted to give the desired resonant frequency in RX and RY.

There are limits to how much the tilt-resonances can be lowered in frequency, such spring-antispring systems are prone to hysteresis and collapse when the quality factor approaches 0.5~\cite{Saulson94.RSI}. Ellipticity in the suspension fibre will also make one axis stiffer than the other. However, for the parameters suggested here, neither of these issues are expected to be significant and there is still margin to reduce the total suspended mass (and as such the fibre diameter), resulting in lower elastic (tilt) resonant frequencies.

Tuning the resonant frequency down, or applying a gravitational anti-spring, `concentrates' the elastic loss, reducing the resulting quality factor like $\omega_{\rm el}^2/\omega_{\rm RX}^2$~\cite{Harms17.CQG}. While this might initially seem bad, the inertial-equivalent motion (or the thermal-noise torque) is only dependent on the loss, regardless of the resonant frequency, and in this case the loss is due to anelasticity in the bending of the suspension wire. 

Since the sensitivity of an inertial isolation system is determined by the free response of the reference mass, the thermal noise is given by
\begin{align}
\sqrt{S_{\rm rot}(\omega)} & = \dfrac{\sqrt{S_{\rm torq}(\omega)}}{\omega^2 I_{\rm RX}}.
\label{eq:thRot}
\end{align}
The total suspended mass and the fibre diameter may be adjusted to achieve a constant fibre stress such that $m \propto r^2$. Combining main text Eqs.~1 and 2, we find that the thermal-noise torque is proportional to $m^{\frac{3}{4}}$, and with supplemental Eq.~\ref{eq:thRot}, the resulting angular motion is proportional to $m^{-\frac{1}{4}}$, as long as the effective radius does not change with mass. 

This weak dependence of thermal noise on the suspended mass means that technical considerations can dictate the design, at the factor of a few level, rather than the elastic resonant frequency or the final quality factor of the RX and RY resonances. The size of the mass, $R$, is chosen such that the rotational thermal noise (barely) limits the final longitudinal inertial sensitivity. It can be readily shown that a steel or tungsten wire would have approximately two orders of magnitude worse performance thermal noise performance, and as such unable to produce satisfactory inertial isolation in this 6D system.

While the coupled equations of motion are important for the tilting modes, the fundamental horizontal translation resonances are nearly unaffected by the bottom hinge and material stiffness, and can be well approximated by a simple pendulum. Z and RZ are essentially independent of other degrees of freedom and the fundamental resonant frequencies can be, respectively, determined by the (effective) extension of the blade spring, $ \Delta z$, and the torsional stiffness of the suspension wire
\begin{align*}
\omega_{\rm X, Y}^2 & = \frac{g}{L},            \\
\omega_{\rm Z}^2 & = \frac{g}{\Delta z},        \\
\omega_{\rm RZ}^2 & = \frac{S J}{L I_{\rm RZ}},
\end{align*}
where $J = \frac{\pi r^4}{2}$ is the second moment of area of the cylindrical fibre along it's long-axis.

\begin{figure}
  \centering
  \includegraphics[width=\columnwidth]{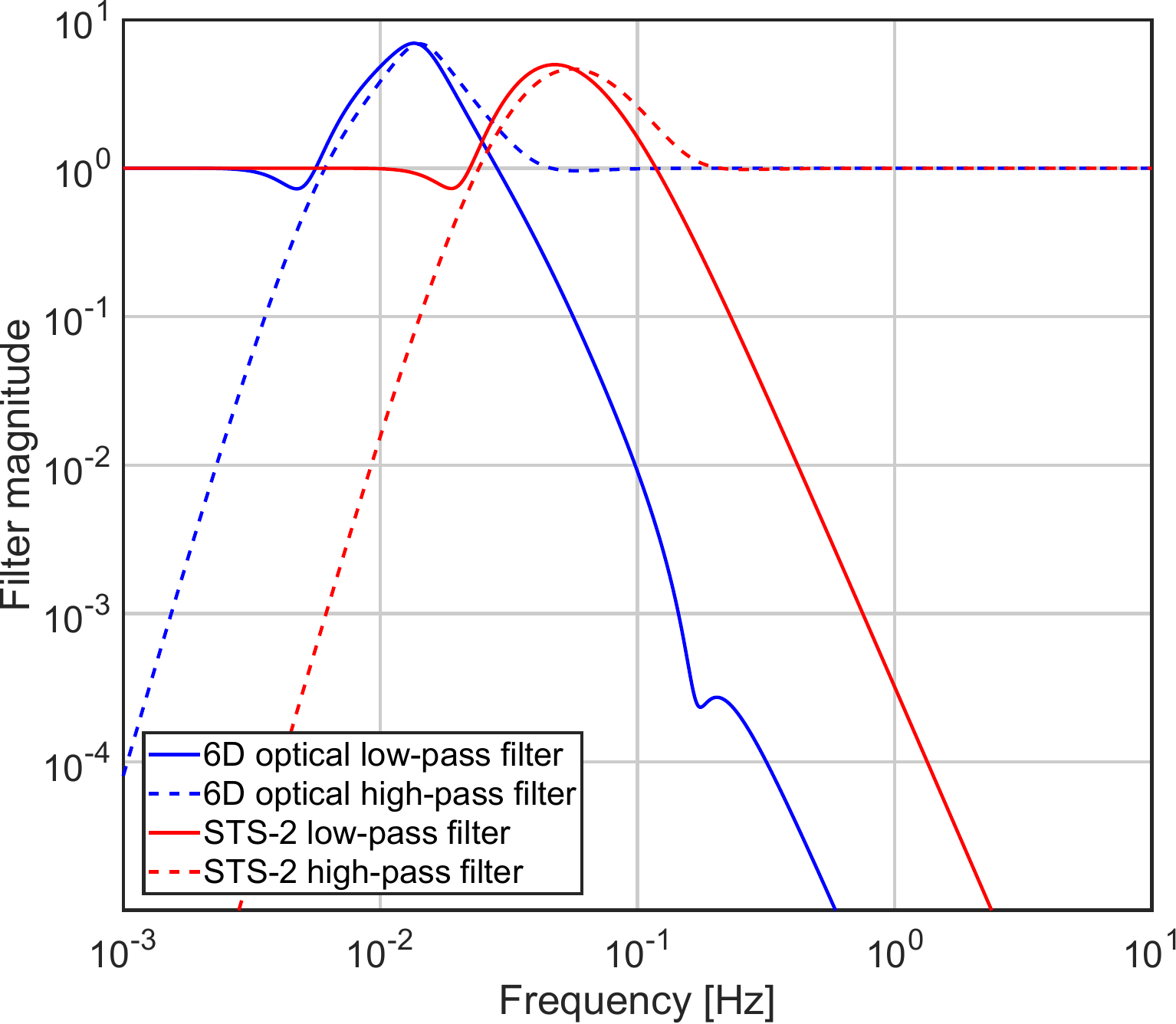}
  \caption{Blend filters for 6D and STS-2 seismometers. Signals from position sensors are used below the blending frequency for feedback control, thereby coupling the platform to the ground.} 
  \label{fig:blend}
\end{figure}

\section{Sensor blending filters}
%\textit{Sensor blending filters} --- 
At low frequencies the sensitivity of LIGO's inertial sensors degrades as $\sim 1/f^{4.5} - 1/f^{5}$ as shown in main-text Fig.~2 (right). In order to avoid contamination of the optical bench motion with this noise, LIGO high passes signals from inertial sensors below a particular frequency. Instead, signals from position sensors that measure the relative motion between the ground and the bench with high precision are used. The sensor blending frequency is determined by the combination of ground-motion injection and seismometer noise, and for both the STS-2 and 6D optical seismometers, it is chosen to minimize the RMS velocity of the isolated platform as a result of the combination of these terms. It is possible to lower the blend frequency, improving performance at 0.1\,Hz, at the expense of substantially increased low-frequency motion.

Fig.~\ref{fig:blend} shows blending filters used to calculate the residual motion of the optical bench shown in main-text Fig.~3 with the 6D seismometer (10\,mHz blend) and STS-2 seismometers (40\,mHz blend). In each case the low- and high-pass filters are complementary (the sum is equal to one).

%%%%%%%%%%%%%%%%%%%%%%%%%%%%% Bibliography %%%%%%%%%%%%%%%%%%%%%%%%%%%%%%

% Commenting out bibliography for ArXiv submission
% \bibliographystyle{apsrev}
% \bibliography{paper}

\end{document}